\documentclass[preprint,aps]{revtex4}
\usepackage{graphicx}
\usepackage{dcolumn}
\usepackage{amsmath}
\begin{document}
\baselineskip 24 true pt   
\title{Polaronic Heat Capacity in The Anderson - Hasegawa Model }

\author{Manidipa Mitra, P. A. Sreeram and Sushanta Dattagupta}
\affiliation{S. N. Bose National Centre for Basic Sciences, JD Block, 
Sector III, Salt Lake City, Kolkata-700098, India}
\email{mmitra@bose.res.in, sreeram@bose.res.in, sdgupta@bose.res.in}
\begin{abstract}

        An exact treatment of the Anderson - Hasegawa two - site model,
        incorporating the presence of superexchange and polarons, is used
        to compute the heat capacity. The calculated results point to the
        dominance of the lattice contribution, especially in the ferromagnetic
        regime. This behavior is in qualitative agreement with experimental
        findings.
~

\noindent


\end{abstract}
\maketitle

	The Anderson-Hasegawa (AH) model \cite{and} is a two-site realization of the basic
idea of double - exchange (DE) proposed by Zener \cite{zen} almost fifty years ago. In the DE scenario
a localized spin is visualized to be strongly `Hund's rule' coupled to an
itinerant spin at the same site governed by strength $J_H$, whereas the 
itinerant spin can tunnel from site to site accompanied by a `hopping
integral' $t$. Because of large $J_H$ the itinerant spin is polarized 
along the localized spin, and as it hops to a neighboring site, it carries
with it the memory of its spin polarization, thereby polarizing the
neighboring local spin as well. Thus transport is correlated with spin 
ordering of localized moments, leading to concomitant metal-insulator 
transition and magnetic ordering. 

Since its inception the DE concept has undergone several extensions including 
a superexchange process yielding antiferromagnetic coupling between localized
moments, as well as polaronic modification of hopping. Indeed polaronic 
contributions are considered to be quite important for thermodynamic 
properties of a doped magnetic system e.g., $La_{1-x}X_xMnO_3$, ($X$ = 
$Ba$, $Ca$, $Sr$ etc). 
Both thermodynamic and transport phenomena in manganites suggest 
the importance of polaron formation and the consequent localization of charge 
carriers \cite{millis95}. 
The two - site 
AH model provides an exactly calculable framework in which some of these
ideas can be tested, for evaluating measurable properties of manganites
in the wider context of a lattice. Besides it is important to keep track
of the quantum nature of the localized spin \cite{muller} -- for instance, 
$Mn^{4+}$ is a spin - 
$\frac{3}{2}$ ion -- even though in much of the DE literature  
the localized moment is viewed as a classical vector. Such an exact quantum
treatment of the two-site AH model incorporating the roles of superexchange
and polarons, and their contributions to phase diagram and heat capacity,
are the subject of this Brief Report. For manganites superexchange interaction
is also influenced by Jahn-Teller (JT)
coupling \cite{fein}, which is however not considered here for the sake
of simplicity.

With the preceding background to the scope and purpose of the present work we
start from the AH model including superexchange for which the Hamiltonian 
can be written as
\begin{equation}
H_{DE}=-t\sum_\tau (c_{1\tau }^{\dagger }c_{2\tau }+h.c.)-J_{H}\sum _{i=1}^{2}
\vec{S_{i}}.\vec{\sigma _{i}} + J \vec{S_1}.\vec{S_2}.
\label{eq1}
\end{equation}
Here $c_{i\tau}^{\dagger}$($c_{i\tau}$) is the creation(annihilation) operator
of the itinerant electron at site $i$ having spin projection $\tau$, 
$\vec S_i$ is 
the localized spin, $\vec \sigma_i$ is the itinerant spin at site $i$ and
$J$ is the strength of the superexchange interaction between neighboring sites.
For our case we consider $\mid \vec S_i \mid = S$, i.e. the localized spins on all sites are taken to have the same value.

From the {\it exact} eigenvalue of $H_{DE}$ we may take the large $J_H$ 
limit by expanding upto $O(1/J_H)$ and write an {\it effective}
Hamiltonian for the two - site one electron case as \cite{mitra}
\begin{equation}
H_{eff} = -t \frac{(S_0 + \frac{1}{2})}{2S + 1} (c_1^\dagger c_2 + h.c)
+ J \vec S_1 . \vec S_2 + \Delta E_J (\hat n_1 + \hat n_2).
\label{eq2}
\end{equation} 
Here $S_0$ is the magnitude of the total spin (localized plus itinerant)
given by $\mid \vec S_1 + \vec S_2 + \vec \sigma \mid$, and 
\begin{equation}
\Delta E_J = \frac{J}{2} {\frac {2S - \bar {S^\prime}}{2S + 1}}
(\bar S^\prime +1),
\label{eq3}
\end{equation}
where $\bar S^\prime = S_0 - 1/2 $.
The first term in Eq. (\ref{eq2}) is the one obtained by Anderson-Hasegawa 
when the localized spin is treated quantum mechanically. The third term,
represented by the number operators
$\hat n_{1(2)}$ for the itinerant electron, modifies the double-exchange
mechanism in the presence of the superexchange interaction given by the
second term in Eq. (\ref{eq2}). This on-site term, proportional to $\Delta E_J$,
we should emphasize, is hitherto not widely considered in the literature,
and is a direct consequence of the quantum nature of the localized spin. The
spin index $\tau$ has been omitted from Eq. (\ref{eq2}), for the sake of 
brevity, as the spin moment of the itinerant electron in any case is parallel
to the localized moment, in the $J_H \longrightarrow \infty$ limit.

We now turn our attention to the polaronic effects.
The minimal model which reflects lattice
carrier interaction on the double-exchange can be introduced by dovetailing
the Holstein mechanism on the Anderson-Hasegawa Hamiltonian. Therefore, 
in the limit of large Hund's rule
coupling, we may write a two site, single polaron, Anderson-Hasegawa-Holstein Hamiltonian from Eq. (\ref{eq2}) as,
\begin{eqnarray}
H &=& H_{eff}
+ g_1 \omega_0  \sum_{i=1}^2  n_{i} (b_i + b_i^{\dag}) 
+ g_2 \omega_0   \left[ n_{1} (b_{2}
+ b_{2}^{\dag}) +n_{2} (b_1 + b_1^\dagger)\right]
+  \omega_0 \sum_{i=1}^2  b_i^{\dag} b_i, 
\label{qdh-1}
\end{eqnarray} 
where, $g_{1}(g_{2})$ denotes the on-site (intersite) electron-phonon 
coupling strength.  
Note that we have considered a single phonon mode for interatomic 
vibrations of frequency $\omega_0$ for which $b_i$ and $b_{i}^{\dag}$ are the 
annihilation and creation operators. The Hamiltonian (\ref{qdh-1}), without
the term $\Delta E_J$, has been the subject of exact 
analytical study for
$S=\infty$ and $S=\frac{1}{2}$ cases and a numerical solution for the
$S=\frac{3}{2}$ case \cite{capo}.

We separate out the in-phase mode and the out-of-phase mode by introducing new 
phonon operators $a=~(b_1+b_2)/ \sqrt 2$ and $d=~(b_1-b_2)/\sqrt 2 $ in
the Hamiltonian. 
The in-phase mode does not couple to the
electronic degrees of freedom whereas the out-of-phase mode does, leading
to an effective electron - phonon Hamiltonian $H_d$, given by,
\begin{equation}
H_d = \omega_0 d^\dagger d + \Delta E_J\sum_{i=1}^2 n_i - t\left(\frac{S_0+\frac{1}{2}}{2S+1}\right) (c_1^\dagger c_2 + h.c.) + g_-\omega_0(n_1-n_2)(d+d^\dagger) 
+J \vec S_1.\vec S_2,
\label{eq5}
\end{equation}
where $g_{-}=(g_1-g_2)/\sqrt 2$.
Following \cite{jayee} we use a Modified Lang-Firsov (MLF) transformation  
and obtain,
\begin{equation}
\tilde{H_d} =  e^R H_d e^{-R},
\label{eq6}
\end{equation}
where $R =\lambda (n_1-n_2) ( d^{\dag}-d)$, $\lambda$ being a
variational parameter related to the displacement of the $d$ oscillator. 
The basis set is given by 
$|\pm,N \rangle = \frac{1}{\sqrt 2} (c_{1}^{\dag} \pm  c_{2}^{\dag})$
$|0\rangle_e  |N\rangle$,    
where $|+\rangle$ and $|-\rangle$ are the bonding and the antibonding 
electronic states and $|N\rangle$ denotes the $N$th excited oscillator  
state within the MLF phonon basis. 
The diagonal part of the Hamiltonian $\tilde{H_d}$ in the chosen basis 
is treated as the unperturbed Hamiltonian ($H_0$) 
and the remaining part of the Hamiltonian $H_{1}= \tilde{H_{d}}-H_0$,  
as the perturbation. 

The unperturbed ground state is the $|+\rangle|0\rangle$ state 
and the unperturbed energy, $ E_0^{(0)}=\epsilon_p - t_{eff} + 
J \vec S_1.\vec S_2 $. 
Where $\epsilon_p = \Delta E_J - \omega_0 ( 2 g_{-} - \lambda) \lambda$ 
and $t_{eff}=t~\frac{S_0 + \frac{1}{2}}{2 S + 1} ~\exp{(-2\lambda^2)}$. 
However, in the exact quantum limit of core spins, 
for given values of $g_-$ and $J$, 
$E_0^{(0)}$ can have four values corresponding to ferromagnetic (FM), 
canted 1 (CA1), canted 2 (CA2) and antiferromagnetic (AFM)
orientation of the two spins for 
$\mid \vec S_{12} \mid = \mid \vec S_1+ \vec S_2 \mid = 3,2,1,0$ 
respectively. Th parameter $\lambda$ is calculated by minimizing the 
unperturbed ground state energy \cite{jayee}.

We have evaluated the perturbation correction to the energy upto the sixth
order and the wave function upto the fifth order. The 
convergence of the perturbation series is very good for $t/\omega_0 \le 1$.
Further, to study the effect of an 
external magnetic field ($\vec h$) we include a term 
$- \tilde g \mu_B (\vec S_1 + \vec S_2). \vec h$ to the Hamiltonian in equation
(\ref{qdh-1}), $\tilde g$ being the Lande g factor. We assume
that the external magnetic field is along the direction of $\vec S_{12}$ and
is expressed in units of $\mu_{eff}(=\tilde g \mu_B)$=1. 

The ferromagnetic (FM) and antiferromagnetic (AFM) orders are related to
$S_{12}(=\mid \vec S_1 + \vec S_2 \mid)=$ 3 and 0,
whereas $S_{12}=$ 2 and 1 are referred to as
canted 1 (CA1) and canted 2 (CA2) states respectively. The Fig. 1 shows the 
phase diagram for the four possible spin orders for our
system, in the $g_{-}$ vs $J$ plane. 
To study the polaronic character one calculates  
the static correlation functions $\langle n_1 u_{1}\rangle_{0}$ and
$\langle n_1 u_{2}\rangle_{0}$,
where $u_1$ and $u_2$ are the lattice deformations at sites 1 and 2
respectively, produced by an electron at site 1 \cite{jayee}. 
The locations of the large polaron region (A) and the
small polaron region (B) are indicated in the $g_{-}$ vs $J$ phase digram (Fig. 1). Different ground states, required for our calculation below, can be located
from the phase diagram in Fig. 1. As our phase diagrams are very similar
to the ones recently presented by Capone and Ciuchi \cite{capo} we henceforth
focus only on our new results for the heat capacity.

As mentioned earlier our main emphasis in this Report is on heat capacity
based on Eq. (\ref{eq5}). Recently, there have been many specific heat 
measurements of the colossal magnetoresistance (CMR) manganites at low 
temperatures with and 
without an external magnetic field \cite{cv, akr, kg, ghiv}. According to experiments the specific heat $C_V$ has contributions from conduction 
electrons, lattice  and spin waves. The low temperature data
\cite{akr,ghiv}, of many CMR materials, show a temperature dependence of 
the form
$C_V = \gamma T + \beta T^3 + \delta T^{3/2}$, here $\gamma$, $\beta $ and 
$\delta$ are constants. The term $\gamma$ arises from charge 
carriers and it is proportional to the density of states at the Fermi level 
and $\beta T^3$ is 
associated with the lattice contribution,  $\beta$ being related to the
Debye temperature. 
The term $\delta T^{3/2}$ gives the spin wave contribution, where 
the coefficient $\delta$ governs the spin wave stiffness. 
Okuda et al \cite{cv} have estimated the electronic specific heat for
$La_{1-x}Sr_xMnO_3$ in the ferromagnetic regime and concluded that the carrier
mass-renormalization near the metal-insulator transition at $x=0.16$ is minimal.
They have also observed a decrease in the low temperature $C_V$ in the
presence of a magnetic field. 
Motivated by these observations, we have carried
out a calculation of the specific heat, based on
the partition function of the system which, from a cumulant
expansion upto the 2nd order, is given by \cite{sd},

\begin{equation}
Z(\beta) = Z_0(\beta) exp{(\int^{\beta}_{0} d\beta^\prime
\int^{\beta^{\prime}}_{0} d\beta^{\prime \prime}\langle \tilde H_1(\beta^\prime)\tilde H_1(\beta^{\prime \prime})\rangle)} ,
\end{equation}
where $Z_0(\beta) = Tr (e^{-\beta H_0})$ ; 
$\tilde H_{1} (\beta) = e^{\beta H_0} H_1 e^{-\beta H_0}$, and 
$\beta = \frac{1}{K_B T}$. The expression $\langle \rangle$ denotes the 
usual canonical averaging. The specific heat is then calculated (in arbitrary
units) from the well known relation:

\begin{equation}
C_V=-\frac{d}{dT}(\frac{d}{d\beta}lnZ(\beta)).
\end{equation}
In the low temperature regime only the zero-and one-phonon states
contribute. 

As the specific heat has a bearing on fundamental properties of CMR
materials it is important to address whether the core spins
should be treated classically ($S \rightarrow \infty $) or quantum
mechanically ($ S = \frac{3}{2}$) for its theoretical estimation.
The difference in the quantum and
classical cases for specific heat, as far as the core spins are concerned,
is exemplified in Fig. 2(a) and Fig. 2(b) for FM and AFM cases respectively.
Temperature is expressed as $\widetilde T= k_B T \omega_0$.
The quantum case only allows for discrete
values of the relative angle between two core spins, while in the classical
case the angle varies continuously. The quantum results evidently yield the correct zero temperature limit.

In the two-site
single polaron model we do not have any scope to vary the carrier 
concentration and
probe different magnetic states. But we can identify the FM and AFM states in 
$g_{-}$ vs $J$
phase diagram, in which the FM state is stable for lower $J$ values 
and the AFM ground state is obtained for larger $J$ values. 
In Fig. 3 we show the $C_V/\widetilde T$ vs $\widetilde T^2$ curves for the FM and AFM cases. 
It is evident
that in the FM case $T^3$ behavior of specific heat is more pronounced which is 
in qualitative agreement with the results of \cite{kg, akr}.
As the stiffness coefficient $\delta $ is proportional to $J$ \cite{kittel},
the spin wave contribution is not significant 
in FM limits. However in the  
AFM limit ($J=0.2$) the variation of $C_V/\widetilde T$ deviates from the $T^3$ law and
the spin wave contribution is non negligible. These findings are in qualitative agreement
with the measurements of Smolyaninova et al \cite{kg}. Moreover the magnitude
of $C_V/\widetilde T$ for the AFM limit is higher than that in 
the $FM$ limit, as in Smolyaninova et al\cite{kg}. The suppression of $C_V$
in the FM regime can be intuitively ascribed to the absence of spin wave
fluctuations, as mentioned earlier. However, while calculating heat capacity
in $S \rightarrow \infty$ limit it can be shown that spin wave contribution
(i.e. $C_V \propto {\widetilde T}^{3/2}$ ) is dominant in FM case. This is due to averaging
over all possible relative orientations of core spins.
  
We show in Fig. 4 the variation of the specific 
heat in the low temperature region in the FM state with zero and one phonon 
states. With application of an external magnetic field $\vec h$, 
$C_V$ takes lower values than for $\vec h= 0$ which is expected, as the average 
energy decreases with application of $\vec h$ in the FM state. 
For CA1($\mid \vec S_{12}\mid=2$), CA2($\mid \vec S_{12}\mid=1$) and AFM 
($\mid \vec S_{12} \mid = 0$) states the external magnetic field will tend 
to align the
core spins to ferromagnetic order($\mid \vec S_{12}\mid = 3$). For CA1, CA2 and 
AFM states at low field and low temperatures
it can be shown from the present calculation that $C_V$ increases
from the $\vec h=0$ limit as long as $\vec h$ does not shift 
$\mid \vec S_{12}\mid$ to higher values. 
For larger $\vec h$, as the ground state changes from lower 
$\mid \vec S_{12}\mid $ to  higher 
ones, $C_V$ decreases in the low temperature region. 
For CMR materials, there are
some reports on measurements of field dependence of $C_V$ in the FM state 
\cite{cv} and also in the half doped case \cite{kg}. It was found that for 
low doping regions, $C_V$ decreases with an increasing 
magnetic field \cite{cv}. However the half doped material showed $C_V/T$ as independent
of applied field \cite{kg}. The present calculation of the external magnetic 
field dependence of $C_V$ qualitatively agrees with these experimental findings 
in the FM limit.
  
In conclusion, the present calculation of $C_V$ using an exactly solvable
model reveals
some of the important features of the double exchange polaronic system.
The discreteness associated with the effective
hopping as a result of the quantum nature of the local spin was shown to have a
significant consequence for thermodynamic properties. 
As analytic calculations of the heat capacity for CMR material to fit 
experimental results are not starightforward, because of the involvement 
of several parameters,
the present calculation for a simplified model 
indeed serves an important role in indicating general trends. Further, a comparison of the computed $C_V$ with measured values underscores the importance
of the quantum nature of the local spin, a fact often ignored in the current
CMR literature.

We are grateful to Sashi Satpathy for discussion on the polaronic mechanism.
SD wishes to thank S. D. Mahanti for generating interest in manganites.

\newpage

\vskip 1.8in

{\bf Figure Captions : }
\vskip 0.5cm

\noindent 

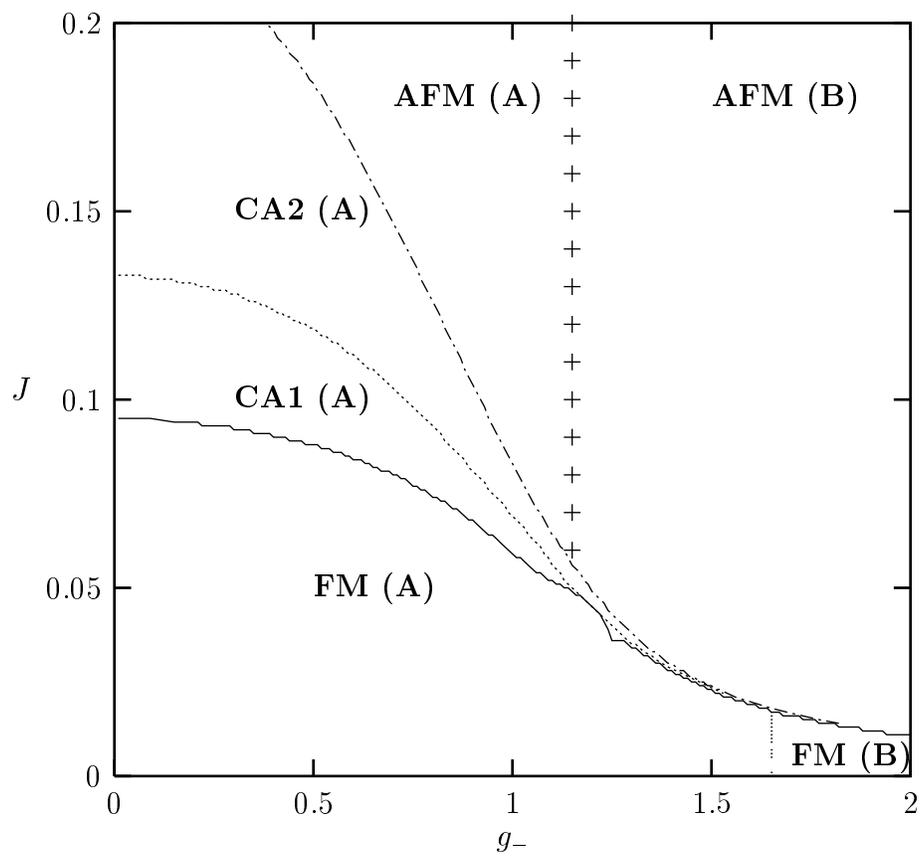
FIG. 1. The $g_{-}$ vs $J$ phase diagram ($\vec h = 0$) for 
$\mid \vec S_1 \mid =
\mid \vec S_2 \mid = \frac{3}{2}$, $t=1$ and $\omega_0=1$. 
({\bf A}) and ({\bf B}) denote
large polaron and small polaron region respectively.

\vskip 0.5cm

\noindent

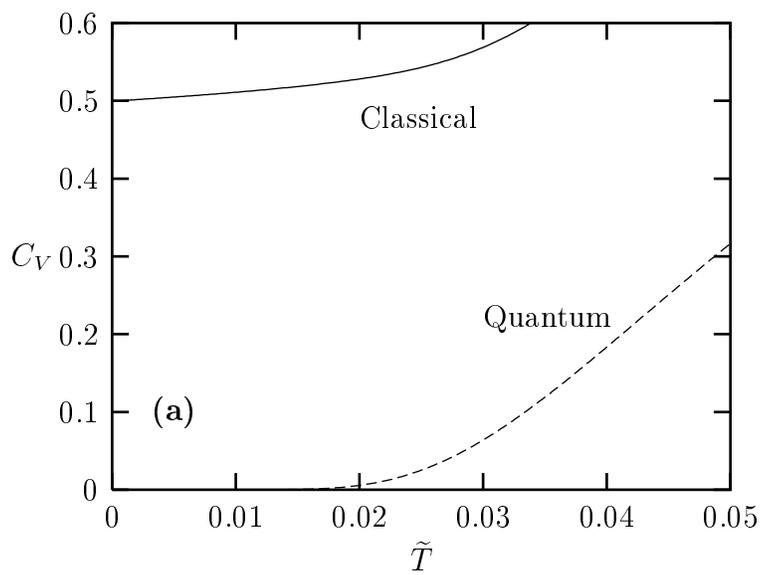
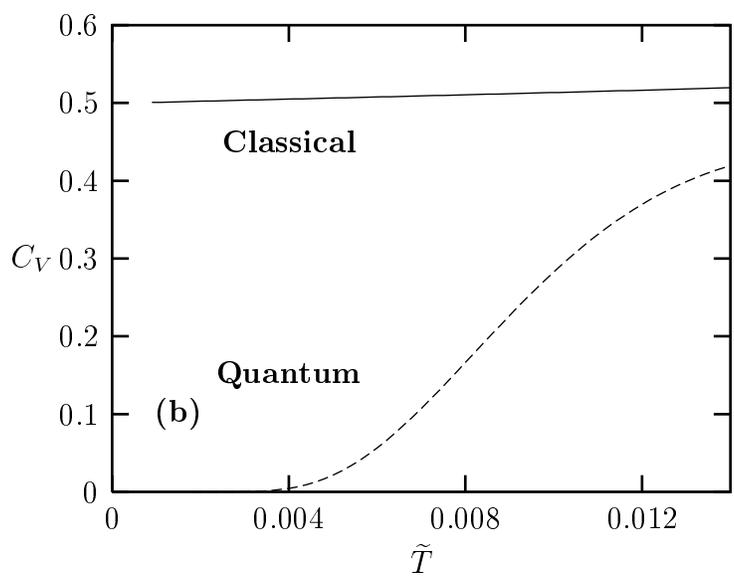
FIG. 2. 
Variations of $C_V$ (in arbitrary units) with temperature 
$\widetilde T(=k_BT \omega_0)$
for $h=0$,
$t=1$, $\omega_0=1$, in classical (solid line) and quantum (dashed line) 
formulations of the core spins for (a) FM ground state($g_-=0.2$, $J=0.02$)
and (b) AFM ground state ($g_-=0.6$, $J=0.2$).

\noindent

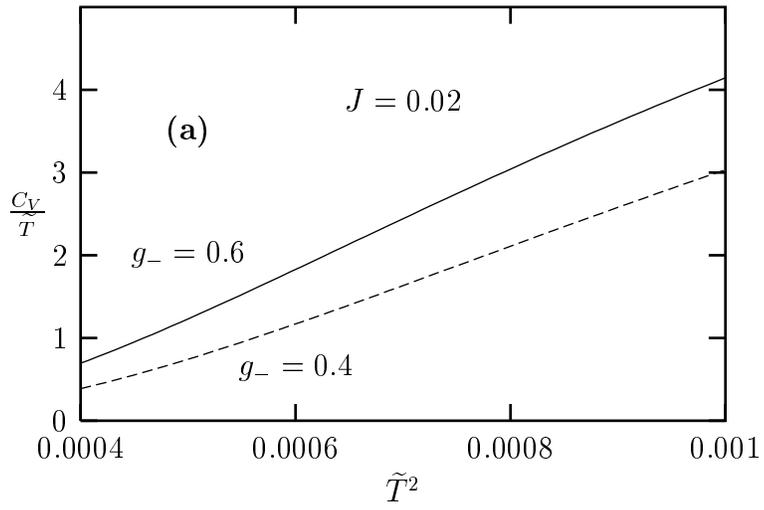
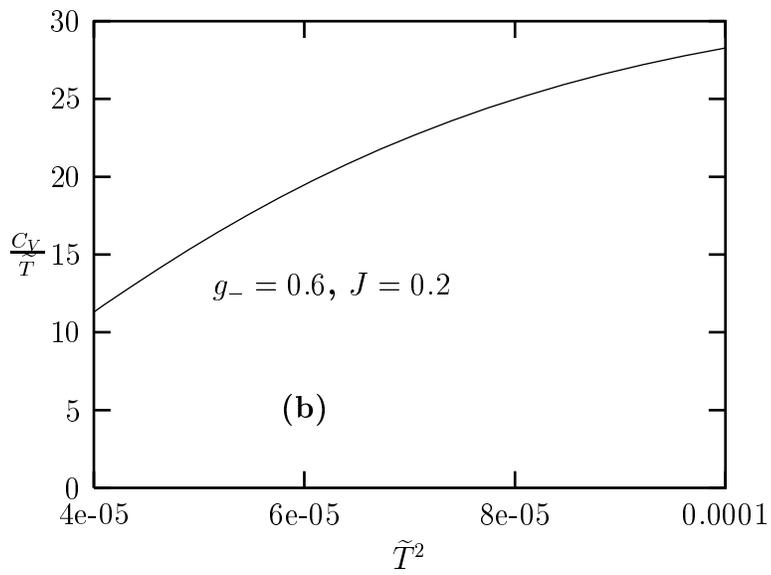
FIG. 3.
$C_V/\widetilde T$ vs $\widetilde T^2$ for $h=0$,
$t=1$, $\omega_0=1$ in (a) FM ground state and (b) AFM ground state
for facilitating comparison with experiments. Here $C_V$ is in arbitrary unit 
and $\widetilde T = k_BT \omega_0$. The results of Fig. 3(a) are in qualitative agreement with
Fig. 2 of Hamilton et al \cite{akr}. 

\noindent

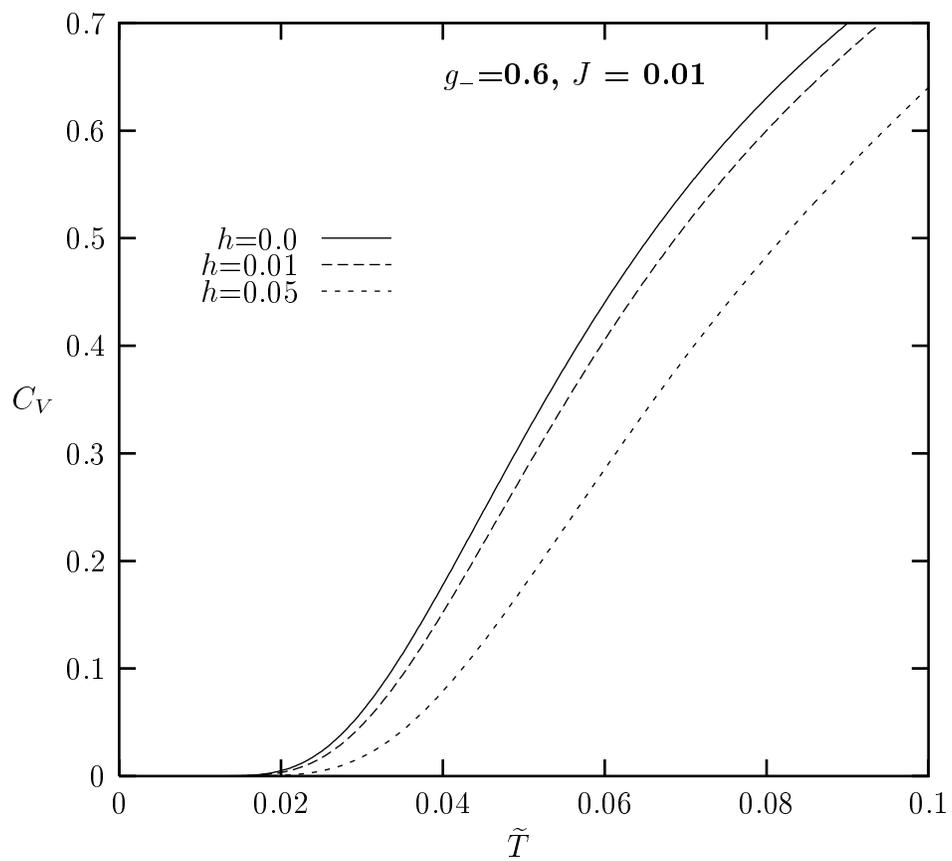
FIG. 4. Variations of $C_V$ (in arbitrary units) for $g_-=0.6$, $J=0.01$ and
$t=1$, $\omega_0=1$, for different values of the magnetic field 
$h=$ 0, 0.01, 0.05, which exemplify the magnetic field dependence of 
$C_V$.
\end{document}